\documentstyle[prl,aps,preprint]{revtex}
\tighten
\begin{document}
\draft
\title{Roughness of Crack Interfaces in Two-Dimensional
       \protect\\Beam Lattices}
\author{Bj{\o}rn Skjetne and T{o}rbj{\o}rn Helle}
\address{Department of Chemical Engineering, \protect\\
         Norwegian University of Science and Technology, \protect\\
         N-7491 Trondheim, Norway}
\author{Alex Hansen}
\address{Institute of Physics, \protect\\
         Norwegian University of Science and Technology, \protect\\
         N-7491 Trondheim, Norway}
\date{\today}
\maketitle
\begin{abstract}
The roughness of crack interfaces is reported in 
quasistatic fracture, using an elastic network 
of beams with random breaking thresholds. For strong
disorders we obtain $\zeta=0.86(3)$ for the roughness 
exponent, a result which is very different from the
minimum energy surface exponent, i.e.,
$\zeta=\case{2}{3}$. A cross-over to lower values is 
observed as the disorder is reduced, the exponent 
in these cases being strongly dependent on the disorder.
\end{abstract}
\narrowtext
Phenomena associated with fracture are a central theme 
in materials science, with great importance in a wide 
range of technological applications. In recent years it 
has also been the subject of much attention in the 
statistical physics community, resulting in methods which 
model materials in terms of disordered rather than 
continuous media~\cite{sta}. This approach has 
drawn attention to certain features apparently sharing 
a common basis with other, seemingly unrelated, problems
showing critical behaviour. For instance, the scaling of 
interfaces which characterize deposition and growth 
processes, or propagation in substrates with a random 
structure, have been found to obey non-trivial laws~\cite{bar}. 
This is also the case of certain equilibrium phenomena, 
where an interface is obtained as a result of interactions 
with a surrounding random medium. In the directed polymer 
problem, for instance, $\zeta=\case{2}{3}$ is obtained 
for the roughness exponent of the minimum energy path in a 
two-dimensional embedding medium~\cite{kar}. 

Furthermore, in numerical simulations with the random 
fuse model~\cite{arc}, the roughness exponent of the
interface which characterizes electrical breakdown of a 
conducting network is found to be $\zeta=0.70(7)$. This
is close to the value for the minimum energy 
path,~$\zeta=\case{2}{3}$~\cite{kar}, and there have been
speculations that they indeed are equal~\cite{rai}.
However, in three dimensions the fuse model exponent 
seems higher than the minimum energy surface 
exponent~\cite{bat}. The two dimensional fuse network 
result~\cite{han} also agrees with experimental results 
in two dimensions~\cite{poi}. However, experimental 
results in three dimensions suggest a much higher 
value,~$\zeta=0.8$, than the fuse model 
gives,~$\zeta=0.62(5)$~\cite{bat}.

A question related to 
whether or not brittle fracture falls within this class 
of problems, however, is how well the random fuse model 
actually describes fracture processes. In the fuse model 
each element has a single degree of freedom, i.e., the 
voltage difference between neighbouring nodes. The 
interface obtained, which is characteristic of electrical 
breakdown rather than a physical crack, nevertheless 
provides valuable information on the interplay between 
quenched disorder and the current distribution. As
opposed to vector fracture, where the elastic elements
each have three degrees of freedom, the random fuse model 
is thus referred to as describing scalar fracture.
It is due to the analogy between Ohm's law and Hooke's 
law that the electrical problem has been regarded as 
similar to its elastic counterpart.

In this Letter, we report the results of computer 
simulations using the elastic beam model~\cite{rou,her} 
which has previously been used to study the
scaling properties of forces and displacements in brittle 
fracture. Furthermore, we address
the universality issue by using two different types of 
distribution with a wide range of disorders.

The beam model in two dimensions may be defined as a 
regular square lattice of size $L\times L$, where the spacing 
is unity, and each node in the horizontal and vertical 
in-plane directions is connected to its nearest 
neighbours by elastic beams. A given beam is then 
soldered to other beams in such a way that, upon 
subsequent displacement of neighbouring nodes, the angle 
between beams remains the same as in the original 
underlying square lattice. The three possible degrees 
of freedom, i.e., translations in the horizontal and 
vertical directions and rotations about the axis 
perpendicular to the plane, thus allow for bending 
moments as well as axial elongation and compression. 
The beam is also imagined as having a certain thickness,
providing shear elasticity. 

The forces between neighbouring nodes may be derived by
considering a concentrated end load on an elastic beam with 
no end restraints~\cite{roa}. We define, for notational 
convenience,
\begin{equation}
    p_{j}=\frac{1}{2}\bigl[1-(-1)^{j}\bigr],
           \hspace{5mm}
    q_{j}=\prod_{n=0}^{j-1}(-1)^{n},
           \label{defq}
\end{equation}
which entails an anti-clockwise labeling beginning with 
the beam to the right of $i$. With $\delta z=z_{j}-z_{i}$ 
denoting the displacements, we obtain at node $i$, due to 
the beam which connects $i$ with $j$,
\begin{eqnarray}
    M_{i}^{(j)}\hspace{-0.5mm}&=&
                \frac{1}{\beta+\frac{\gamma}{12}}
                 \bigl[\hspace{0.5mm}\frac{\beta}{\gamma}
                  \delta\theta
                  +\frac{q_{j}}{2}T_{i}^{(j)}
                   -\frac{\theta_{i}}{3}
                    -\frac{\theta_{j}}{6}
                      \hspace{0.5mm}\bigr],
                       \nonumber\\
    V_{i}^{(j)}\hspace{-0.5mm}&=&
                \frac{1}{\beta+\frac{\gamma}{12}}
                 \bigl[\hspace{0.5mm}T_{i}^{(j)}\hspace{-1.0mm}
                 -\hspace{-0.5mm}\frac{q_{j}}{2}
                   \bigl(\theta_{i}+\theta_{j}\bigr)
                    \bigr],
                     \label{cont}\\
    S_{i}^{(j)}\hspace{-0.5mm}&=&
                \frac{1}{\alpha}
                 \bigl[\hspace{0.5mm}T_{i}^{(j+1)}
                  \hspace{0.5mm}\bigr],
                   \nonumber
\end{eqnarray}
where $T_{i}^{(j)}=\delta x+p_{j}(\delta y-\delta x)$,
for the contributions in moment, shear and strain, respectively. 

Prefactors characteristic of the material and its dimensions 
in Eq.~(\ref{cont}) depend on
\begin{eqnarray}
    \alpha=\frac{1}{EA},\hspace{3mm}
     \beta=\frac{1}{GA},\hspace{3mm}
      \gamma=\frac{1}{EI},
       \label{mate}
\end{eqnarray}
where $E$ is Young's modulus, $A$ and $I$ the 
area of the beam section and its moment of inertia about 
the centroidal axis, respectively, and $G$ the shear modulus.

For the sum of forces and moments on each node, we then have
\begin{eqnarray}
    \Sigma_{ix}&=&S_{i}^{(1)}+V_{i}^{(2)}+S_{i}^{(3)}+V_{i}^{(4)},
       \nonumber\\
    \Sigma_{iy}&=&V_{i}^{(1)}+S_{i}^{(2)}+V_{i}^{(3)}+S_{i}^{(4)},
       \label{forc}\\
    \Sigma_{i\theta}&=&\sum_{j=1}^{4}M_{i}^{(j)},
       \nonumber
\end{eqnarray}
the lattice being in equilibrium when, at any point in the
fracture, $\Sigma_{ix}=\Sigma_{iy}=\Sigma_{i\theta}=0$.
Such a configuration is realized when the 
elastic energy, i.e., 
\begin{equation}
    {\cal E}=\frac{1}{2}\sum_{i}\sum_{j=1}^{2}
              \Bigl\{
               \bigl[S_{i}^{(j)}\bigr]^{2}\hspace{-1.0mm}
               +\bigl[V_{i}^{(j)}\bigr]^{2}\hspace{-1.0mm}
                +\bigl[M_{i}^{(j)}\bigr]^{2}
                  \Bigr\},
                   \label{ener}
\end{equation}
is at its minimum. This minimum we obtain via relaxation,
using the conjugate gradient method with a
tolerance in the residual error of $\epsilon=10^{-12}$. 

For a brittle material we assume that each beam is linearly 
elastic up to the breaking threshold. Using $t_{S}$ and 
$t_{M}$ for the strain and bending thresholds 
respectively, the breaking criterion~\cite{her}, inspired 
from Tresca's formula, is given by
\begin{equation}
    \left(\frac{S}{t_{S}}\right)^{2}+
     \frac{|M|}{t_{M}}\geq 1,
      \label{tresca}
\end{equation} 
where 
$|M|={\rm max}(|M_{i}|,|M_{j}|)$ is the largest of the 
bending moments at the two beam ends $i$ and $j$. 

The fracture process is initiated by imposing on the 
lattice an external vertical displacement of unit 
magnitude, i.e., a displacement which at the top row
corresponds to one beam in length. In its initial state,
the lattice now consists of horizontally undeformed
beams and beams which in the vertical direction are 
stretched lengthwise by an amount $1/L$. With an extra 
row at the top there are $L(L-1)$ inner nodes, for 
which any neighbouring beam may be broken, and $L$ nodes 
each at the top and bottom, the positions of which are 
held fixed. This defines the vertical boundary 
conditions.

As for the horizontal direction, previous results 
obtained with the random fuse model have relied on the 
use of periodic boundary conditions. This is a good 
strategy to avoid edge-effects, especially in 
a situation where numerical resources are limited to 
small system sizes. However, when considering fracture 
in a periodic system, the topology is essentially that 
of a plane intersecting a cylinder. We thus need to 
address the problem of how the trace of a sine curve 
affects results obtained for the roughness. To avoid 
this, we instead use open boundary conditions, i.e., 
we adopt the procedure used in Ref.~\cite{mal} of 
subtracting the average vertical drift of the crack 
as it traverses the width of the lattice.

The first beam to break is that for which the sum
of the two ratios is largest, this being the vertically 
oriented beam which has the lowest value of $t_{S}$.

If all threshold values are approximately the same, the 
next beam to break will be one of the nearest lateral 
neighbours since these now carry a larger load than 
other beams in the lattice. The case of no disorder is 
thus one in which the crack propagates horizontally 
from the initial damage, taking the shortest possible 
path to break the lattice apart. This results in a smooth 
interface. 

Introducing disorder in the threshold values, material 
strength is no longer uniformly distributed throughout the 
lattice and consequently the crack will not necessarily 
develop from the initial damage point. Instead microcracks 
and voids form wherever the stress concentration
most exceeds the local strength, i.e., 
wherever Eq.~(\ref{tresca}) dictates that the next beam 
should be broken. Towards the end of the process some 
of these merge into a macroscopic crack, forming a 
sinuous, or rough, interface which is characteristic 
of the disorder in the system.

Hence we have a highly correlated process in which the 
quenched disorder and the non-uniform stress distribution 
combine to determine where the next break will occur while, 
simultaneously, the stress distribution itself 
continually changes as the damage spreads.

To study this, we generate a random number $r$ on the unit 
interval $[0,1]$ and let this represent the cumulative 
threshold distribution. Assigning the threshold values 
according to 
\begin{equation}
    t_{F}=r^{D}, 
       \label{pf}
\end{equation}
the threshold distribution 
approaches that of no disorder when $|D|\rightarrow0$. 
In the fuse model~\cite{han,dea}, several types 
of distribution have been used for the threshold values.
Although at present we restrict ourselves to Eq.~(\ref{pf}),
the two cases $D<0$ and $D>0$ represent widely different 
distributions, i.e., for $D>0$ the distribution is a power 
law with a tail which extends towards weak beams whereas 
for $D<0$ the tail of the distribution extends towards 
strong beams. In both cases we use a wide range of 
disorders between $|D|=\frac{1}{12}$ and $|D|=4$.

The roughness is now obtained for a large number of lattices, 
each of size $L$, the thresholds being re-cast according to 
Eq.~(\ref{pf}) each time a new sample is broken. Generally 
the number of samples 
depend on $L$ as well as, to a lesser degree, on the 
disorder~$D$. Presently lattices of all sizes from $L=4$ up 
to $L=20$ were studied, with sample sizes ranging from 
$N=250000$ in  the
the former case to about $N=1000$ in the latter case. 
For the larger systems we studied, 
typical sample sample sizes are shown in Table~\ref{one}.

Fig.~\ref{one} shows a log-log plot of $W$ as a function of 
$L$ for a range of 
disorders with $D>0$. For all $L$, the interface is seen to
become more rough with increasing disorder. Each curve also 
has a characteristic crossover, beyond which the asymptotic
relationship is that of a straight line, i.e., where 
$W\sim L^{\zeta}$. This feature is seen 
to be disorder dependent, with the onset of asymptotic 
behaviour being deferred to larger $L$ as $D$ increases.
At some point, the crossover becomes difficult to define
before it again reduces to a point well within the range
of the system sizes presently studied. Closely associated
with this behaviour is an even more striking feature, i.e., 
the dependency of $\zeta$ upon the disorder. Specifically, 
with the enumeration scheme used for the disorders in 
Fig.~\ref{one}, we obtain (f)~$\zeta=0.16$ for $D=0.25$ as 
opposed to (k)~$\zeta=0.55$ for $D=0.09$, between which $\zeta$ 
increases monotonously as the disorder decreases, i.e., 
(g)~$\zeta=0.23$, (h)~$\zeta=0.32$, (i)~$\zeta=0.42$ and 
(j)~$\zeta=0.51$. A very pronounced transition is observed 
between (d)~$D=0.5$ and (f)~$D=0.25$, the exponents for 
$D>0.5$ apparently having a constant value of $\zeta\sim0.86$, 
that is, we obtain (a)~$\zeta=0.87$, (b)~$\zeta=0.86$ and 
(c)~$\zeta=0.86$. 

Although values obtained for $\zeta$ with $D<0$, shown in
Fig.~\ref{two}, are seen to be different from those obtained 
with the same $|D|$ when $D>0$, the qualitative features
in this case are the same. Again there is a pronounced 
transition in $\zeta$ from (a)~$\zeta=0.87$ and (b)~$\zeta=0.86$ 
to (e)~$\zeta=0.41$, 
between which the exponent is difficult to determine. With a 
further decrease in disorder the exponent increases 
up towards (k)~$\zeta=0.60$, the intermediate values shown in
Fig.~\ref{two} being (f)~$\zeta=0.48$, (g)~$\zeta=0.51$,
(h)~$\zeta=0.55$, (i)~$\zeta=0.57$ and (j)~$\zeta=0.60$. 

The behaviour of $\zeta$ as a function of the disorder $D$ 
is shown in Fig.~\ref{zod}. Here, estimates for $\zeta$ 
which are difficult to define are also included. Hence,
corresponding to the open circles in Fig.~\ref{one} we use 
(d)~$\zeta=0.89$, based on the data for $L=27$ to $L=100$ 
and (e)~$\zeta=0.31$, based on the four uppermost data points. 
In Fig.~\ref{two} the corresponding estimates are 
(c)~$\zeta=0.88$, based on data for $L=19$ up to $L=63$,
and (d)~$\zeta=0.43$, again based on the four uppermost
data points. As $|D|\rightarrow 0$, the interface becomes 
sufficiently smooth to frequently avoid detection by the 
course-graining of the lattice. Hence an accurate estimate 
for $\zeta$ now depends on the relative occurrence of those
samples which are unusually rough for the given disorder,
implying an excessively large amount of samples for each $L$.
To obtain an estimate nonetheless, we note that the two
sets of exponents for the six lowest values of $|D|$ 
corresponding to $D<0$ and $D>0$, respectively, each very 
nearly lie on a straight line. The intersection between 
the two lines is $D\approx0.04$, with a limiting value of 
$\zeta\approx0.65$ for the exponent. Although this is 
very close to the two-thirds value frequently referred to 
in connection with scalar fracture, the result obtained for 
$D=0.08$ does not significantly alter the $D=0.1$ result,
which is $\zeta=0.60$. Hence, the lines may taper off at 
this value, the limit $|D|\rightarrow0$ possibly representing 
a Laplacian random walk~\cite{lrw} whereby crack 
advancement is governed by local conditions surrounding 
the crack tip.

Recently the role of propagating stress waves during brittle 
fracture has been investigated~\cite{ram}.
In our model the elastic wave 
emitted from a breaking beam would then result in stresses 
exceeding those due to the elastic deformations only, the 
stress enhancement being especially important in the case 
of an imminent burst of failures. Although this feature
is not included in the present quasistatic approach, the 
comparison with experimental results
for~$\zeta$ in two dimensions~\cite{poi} should remain valid,
i.e., Poirer {\it et al.} obtained $\zeta=0.73\pm0.07$ by 
considering a two dimensional stacking of parallel 
collapsible cylinders (drinking straws) while Kertesz
{\it et al.} and Eng{\o}y {\it et al.} obtained
$\zeta\approx0.73$ and $\zeta=0.68\pm0.04$ by studying 
tear lines in (wet) paper and fractures in thin wood plates,
respectively, none of which should generate stress waves
significant enough to modify the result.

To summarize, the main feature of our results is the 
dependency of $\zeta$ upon the disorder, apparently 
contradicting a universal value. Whereas values obtained 
at low disorders vary considerably, however, the more 
or less constant $\zeta$ obtained at moderate and strong 
disorders nevertheless seems to be consistent with a 
universal value of $\zeta\sim0.86$. While thus being 
similar to the experimental results in {\it three} 
dimensions, our results are different from other two 
dimensional results.

\clearpage
\begin{table}
\caption{Typical number $N$ of lattices generated for
         the various system sizes $L$.} 
\begin{tabular}{crrrrrrrrrrrc}
       & \multicolumn{1}{c}{$L$} & \multicolumn{1}{c}{$N$} & &
         \multicolumn{1}{c}{$L$} & \multicolumn{1}{c}{$N$} & &
         \multicolumn{1}{c}{$L$} & \multicolumn{1}{c}{$N$} & &
         \multicolumn{1}{c}{$L$} & \multicolumn{1}{c}{$N$} &
             \\
              \hline
   & 23 &  800   & &   40 & 300   & &    80 & 150   & &   160 & 50 & \\
   & 27 &  600   & &   50 & 250   & &   100 & 100   & &   200 & 30 & \\
   & 32 &  500   & &   63 & 200   & &   125 &  75   & &   250 & 20 & \\
\end{tabular}
\label{numb}
\end{table}
\noindent

\clearpage
\begin{figure}
\caption{The average roughness $W$ as a function of lattice size 
         $L$, for disorders with $D>0$. The enumeration scheme is 
         (a)~top to (k)~bottom, with $|D|$ equal to (a)~4, (b)~2, 
         (c)~1, (d)~0.5, (e)~0.33, (f)~0.25, (g)~0.20, (h)~0.17,
         (i)~0.14, (j)~0.10 and (k)~0.09.}
\label{one}
\end{figure}
\noindent

\begin{figure}
\caption{The average roughness $W$ as a function of lattice size 
         $L$, for disorders with $D<0$. The enumeration scheme is 
         (a)~top to (k)~bottom, with $|D|$ equal to (a)~4, (b)~2, 
         (c)~1.5, (d)~1, (e)~0.50, (f)~0.33, (g)~0.25, (h)~0.20,
         (i)~0.14, (j)~0.10 and (k)~0.08.}
\label{two}
\end{figure}
\noindent

\begin{figure}
\caption{The roughness exponent $\zeta$ as a function of the disorder 
         $D$, with thresholds chosen according to Eq.~(\ref{pf}). 
         Labels and symbols refer to the enumeration scheme used in 
         Fig.~\ref{one} and Fig.~\ref{two}, with ({\large$\star$}) 
         referring to the extrapolated value for $D\approx0$, 
         i.e., $\zeta=0.65$.}
\label{zod}
\end{figure}
\noindent

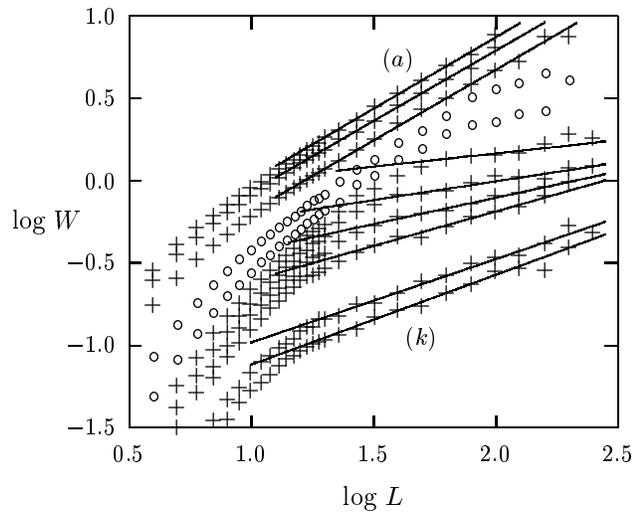
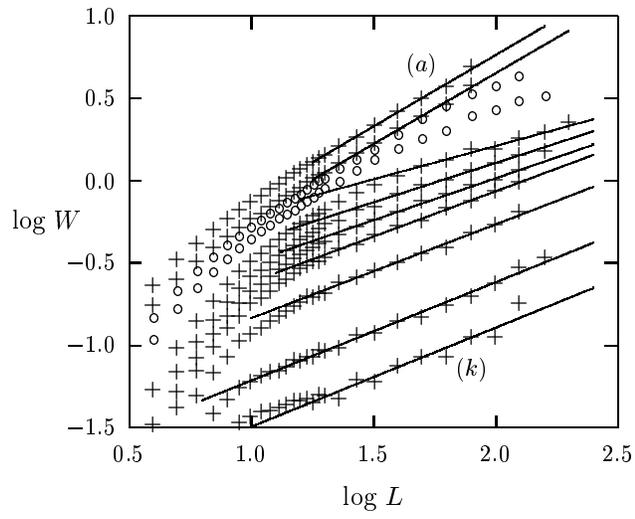
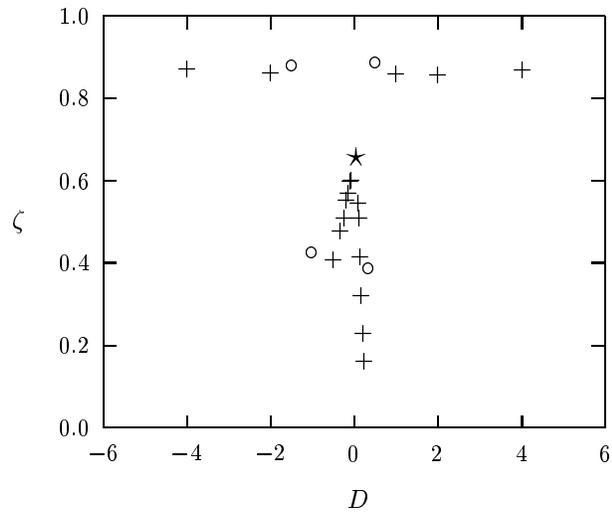
\end{document}